\documentclass[sigconf,natbib=true]{acmart}
\usepackage{xspace}
\usepackage{soul}
\usepackage{url}
\usepackage{graphicx}
\usepackage{amsmath}
\usepackage{amsfonts}
\usepackage{multirow}
\usepackage{amsthm}
\usepackage{booktabs}
\usepackage{epstopdf}
\usepackage{xcolor}
\usepackage[labelformat=simple]{subcaption}
\usepackage{caption}
\usepackage{bbm}
\usepackage{dsfont}
\usepackage[ruled,linesnumbered]{algorithm2e}
\usepackage{setspace}
\usepackage{braket}
\usepackage{tablefootnote}
\usepackage{subcaption}
\usepackage[normalem]{ulem}
\useunder{\uline}{\ul}{}

\newcommand{\ourmethod}{\texttt{NP-FKGC}\xspace}
\graphicspath{{images/}}

\AtBeginDocument{%
  \providecommand\BibTeX{{%
    \normalfont B\kern-0.5em{\scshape i\kern-0.25em b}\kern-0.8em\TeX}}}

\copyrightyear{2023}
\acmYear{2023}
\setcopyright{acmlicensed}\acmConference[SIGIR '23]{Proceedings of the 46th
International ACM SIGIR Conference on Research and Development in
Information Retrieval}{July 23--27, 2023}{Taipei, Taiwan}
\acmBooktitle{Proceedings of the 46th International ACM SIGIR Conference on
Research and Development in Information Retrieval (SIGIR '23), July 23--27,
2023, Taipei, Taiwan}
\acmPrice{15.00}
\acmDOI{10.1145/3539618.3591743}
\acmISBN{978-1-4503-9408-6/23/07}

\begin{document}

\title{Normalizing Flow-based Neural Process for Few-Shot Knowledge Graph Completion}

\author{Linhao Luo}
\email{linhao.luo@monash.edu}
\affiliation{%
  \institution{Monash University}
    \country{Australia}
}

\author{Yuan-Fang Li}
\email{yuanfang.li@monash.edu}
\affiliation{%
  \institution{Monash University}
    \country{Australia}
}

\author{Gholamreza Haffari}
\email{Gholamreza.Haffari@monash.edu}
\affiliation{%
  \institution{Monash University}
    \country{Australia}
}

\author{Shirui Pan}
\authornote{Corresponding author.}
\email{s.pan@griffith.edu.au}
\affiliation{%
  \institution{Griffith University}
  \country{Australia}
 }

\renewcommand{\shortauthors}{Linhao Luo, Yuan-Fang Li, Gholamreza Haffari, and Shirui Pan}

\begin{abstract}
    Knowledge graphs (KGs), as a structured form of knowledge representation, have been widely applied in the real world. Recently, few-shot knowledge graph completion (FKGC), which aims to predict missing facts for unseen relations with few-shot associated facts, has attracted increasing attention from practitioners and researchers. 
    However, existing FKGC methods are based on metric learning or meta-learning, which often suffer from the out-of-distribution and overfitting problems. Meanwhile, they are incompetent at estimating uncertainties  in predictions, which is critically important as model predictions could be very unreliable in few-shot settings. Furthermore, most of them cannot handle complex relations and ignore path information in KGs, which largely limits their performance.
    In this paper, we propose a normalizing flow-based neural process for few-shot knowledge graph completion (\ourmethod). Specifically, we unify normalizing flows and neural processes to model a complex distribution of KG completion functions. This offers a novel way to predict facts for few-shot relations while estimating the uncertainty.
    Then, we propose a stochastic ManifoldE decoder to incorporate the neural process and handle complex relations in few-shot settings. To further improve performance, we introduce an attentive relation path-based graph neural network to capture path information in KGs. Extensive experiments on three public datasets demonstrate that our method significantly outperforms the existing FKGC methods and achieves state-of-the-art performance. Code is available at \url{https://github.com/RManLuo/NP-FKGC.git}. 
\end{abstract}




\maketitle
\section{Introduction}\label{sec:introduction}

\begin{figure}[tbp]
    \centering
    \includegraphics[width=.98\columnwidth,trim=0 0 0cm 0, clip]{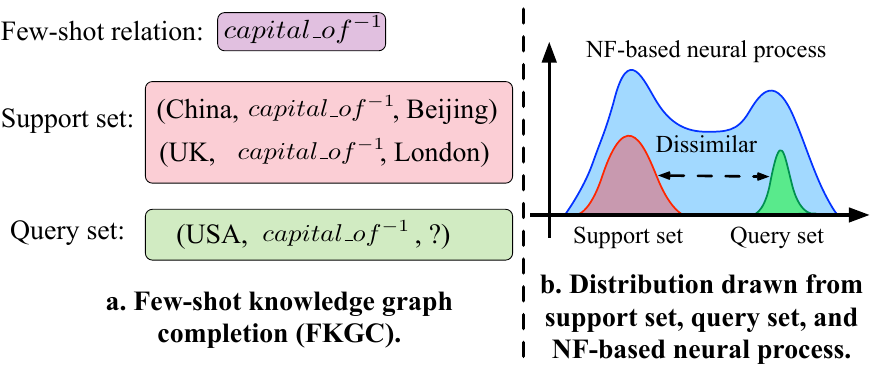}
    \caption{(a.) An example of few-shot knowledge graph completion and (b.) distribution drawn from support set, query set, and our approach.}
    \label{fig:intro}
\end{figure}

Knowledge graphs (KGs) containing enormous facts in a way of relations between entities, i.e., $(head~entity, relation, tail~entity)$, are a structured form of knowledge representation (e.g., \ YAGO \cite{suchanek2007yago}, NELL \cite{carlson2010toward}, and Wikidata \cite{vrandevcic2014wikidata}). These KGs are essential for many applications \cite{ji2021survey}, such as web search \cite{xiong2017explicit}, question answering \cite{zhang2018variational}, and recommender systems \cite{wang2019multi}.
Real-world KGs are often incomplete. Many studies have been proposed to complete missing facts by inferring from existing ones \cite{bordes2013translating,wan2021reasoning,xiong2022ultrahyperbolic,wang2022exploring}. However, these methods require sufficient training facts for each relation, which is not always available. For instance, a large portion of relations in KGs are long-tailed with less than 10 associated facts \cite{xiong2018one}. This largely impairs the performance of existing methods. Thus, it is essential and challenging to complete facts for relations in a few-shot manner.

Few-shot knowledge graph completion (FKGC) aims to design a KG completion function (model) that predicts unseen facts of a given relation in a \textit{query set} with few-shot triples (facts) in a \textit{support set}. For example, as shown in Fig.~\ref{fig:intro}a, given a relation $capital\_of^{-1}$ and two triples in its support set, we aim to predict the fact: ``which city is the capital of USA?'' in the query set, i.e., $(USA, capital\_of^{-1}, ?)$. Since we have two triples in the support set, this can be denoted as a 2-shot FKGC problem.

Existing FKGC methods can be grouped into two categories: metric learning-based methods \cite{xiong2018one,zhang2020few,sheng2020adaptive} and meta-learning-based methods \cite{chen2019meta,niu2021relational}. However, they often suffer from out-of-distribution \cite{huangfew} and overfitting problems \cite{dong2020mamo}.
For example, as shown in Fig. \ref{fig:intro}b, if triples in the query set share dissimilar distribution with the support set, metric learning-based methods would fail to predict unseen facts precisely based on similarities calculated from these two sets. By updating model parameters with the support set, meta-learning-based methods are also prone to get stuck at a local optimum \cite{antonioutrain}. Furthermore, they cannot quantify uncertainties in their predictions, which is crucial for generating reliable predictions under few-shot scenarios \cite{zhanguncertainty,mukherjee2020uncertainty}. Additionally, present approaches fall short of considering path information between entities, which is also essential for the FKGC task \cite{xu2021p,zhu2021neural}.

Neural processes (NPs) \cite{garnelo2018neural} are a family of methods offering a new way to deal with tasks with limited data. NPs are based on the stochastic process, which model a distribution over prediction functions. The distribution is defined by a latent variable model conditioned on limited data. By modeling the distribution, NPs can not only estimate uncertainties over predictions but also generalize to new tasks with few-shot data. As shown in Fig. \ref{fig:intro}b, NP-based methods can define a more general distribution, incorporating both the support set and query set.

\begin{figure}[tbp]
    \centering
    \includegraphics[width=0.8\columnwidth,trim=0 0 0cm 0, clip]{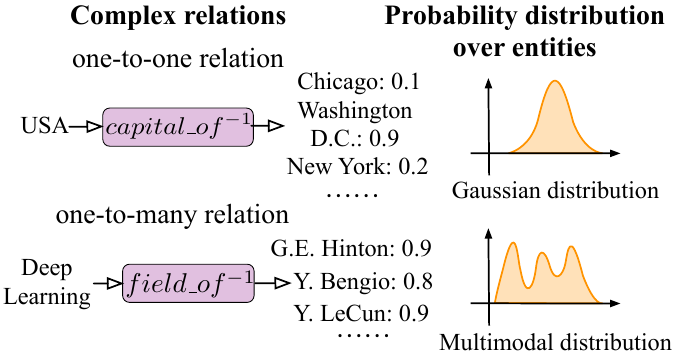}
    \caption{Complex relations and their corresponding probability distributions over entities.}
    \label{fig:comple_rel}
\end{figure}

However, NPs \cite{kim2018attentive,singh2019sequential,cangea2022message} cannot be applied directly to the FKGC task. Conventionally, NPs adopt a Gaussian distribution to model the stochastic process, which cannot handle complex relations in KGs. In Fig. \ref{fig:comple_rel}, we illustrate two types of complex relations (i.e., \textit{one-to-one} and \textit{one-to-many} relations). one-to-one relations are those that, when given a head entity, there is only 1 possible tail entity. For example, the capital of \textit{USA} can only be \textit{Washington D.C.}. Therefore, the tail entity's probability distribution satisfies the Gaussian distribution. For one-to-many relations, multiple tail entities could be linked to the same head entity. For example, all \textit{G.E. Hinton}, \textit{Y. Bengio}, and \textit{Y. LeCun} are in the field of \textit{deep learning}, where the probability distribution could be a multimodal distribution. Besides, a Gaussian distribution would lead to the posterior collapse \cite{razavi2018preventing} and provide meaningless information. Thus, it is inappropriate to use a simple Gaussian distribution to model KG completion functions.

To address the aforementioned challenges, we propose a neural process-based approach for FKGC, named \ourmethod. Specifically, we first adopt the neural process (NP) to define a distribution over KG completion functions. By sampling from the distribution, we can readily obtain a prediction function that is specialized for the given few-shot relation. Then, the NP is extended by a normalizing flow (NF) \cite{papamakarios2021normalizing} to transform the simple Gaussian distribution into a complex target distribution, which is more expressive to handle complex relations and estimate uncertainties. Furthermore, we propose a novel stochastic ManifoldE decoder (SManifoldE) to predict facts in the query set. By marrying the merits of neural process and ManifoldE \cite{xiao2016one}, SManifoldE could handle complex relations with limited observed triples. Last, we propose an attentive relation path-based graph neural network (ARP-GNN) to effectively consider paths for FKGC. Instead of using the Breadth-first search (BFS), we adopt a graph neural network (GNN) to encode path information into entity representations attentively. In this way, \ourmethod can consider both complex relations and path information. Extensive experiments on three public benchmark datasets show that \ourmethod significantly outperforms state-of-the-art methods by a large margin (33.7\%, 43.3\%, and 17.5 in MRR as well as 77.6\%, 41.5\%, and 36.4\% in Hits@1, respectively).  

The main contributions of this paper are summarized as follows:
\begin{itemize}
    \item We propose a neural process-based few-shot knowledge graph completion method (\ourmethod). To the best of our knowledge, this is the first work that applies NPs to the FKGC problem.
    \item We propose a novel normalizing flow-based neural process encoder and a stochastic ManifoldE decoder (SManifoldE) to simultaneously estimate uncertainties and handle complex relations. An attentive relation path-based GNN (ARP-GNN) is also designed to incorporate path information for FKGC.
    \item We conduct extensive experiments on three public benchmark datasets. Experiment results show that \ourmethod can significantly outperform the state-of-the-art methods.
\end{itemize}
\section{Related work}\label{sec:relatedwork}

\subsection{Few-shot Knowledge Graph Completion}
Existing few-shot knowledge graph completion (FKGC) methods can be roughly divided into two groups: metric learning-based and meta-learning-based models. Metric learning-based methods develop a matching network to calculate similarities between triples in the support set and query set. GMatching \cite{xiong2018one} is the first work in FKGC, which proposes a neighbor encoder and a LSTM matching network to measure the similarity. FSRL \cite{zhang2020few} extends the GMatching to simultaneously consider multiple support triples with an attention neighbor encoder and a LSTM support set encoder. On the top of FSRL, FAAN \cite{sheng2020adaptive} presents a relation-specific adaptive neighbor encoder. Meanwhile, it adopts a transformer-based encoder to learn representations of triples. To consider uncertainties of KGs, GMUC \cite{zhang2021gaussian} proposes a Gaussian-based metric function that captures uncertain similarities. With the power of GNN \cite{zheng2022rethinking,jin2022multivariate,jin2022neural}, CSR \cite{huangfew} introduces a sub-graph-based matching network with pre-training techniques. P-INT \cite{xu2021p} introduces a path-based matching network. However, it fails to address situations where entities cannot be connected by paths. Besides, the matching network cannot handle complex relations in FKGC. 

Meta-learning-based models aim to quickly update model parameters for an unseen relation. MetaR \cite{chen2019meta} proposes a relation-meta learner to represent few-shot relations and update their representation using the support set. MetaP \cite{jiang2021metap} introduces a meta pattern learning framework to predict new facts. GANA \cite{niu2021relational} integrates meta-learning with TransH \cite{wang2014knowledge} and brings MTransH to handle complex relations. Besides, GANA also introduces a gated and attentive neighbor aggregator to address challenges of sparse neighbors. But, the meta-learning-based methods require a fine-tuning process to update model parameters, which is not efficient enough since it needs to calculate and store gradients. What is more, they suffer from out-of-distribution problems \cite{liu2022beyond,tan2022federated,liu2023good} and fail to quantify uncertainties in their predictions.

\subsection{Neural Processes}
Neural Processes (NPs) \cite{garnelo2018neural} combine the stochastic process and neural networks to define a distribution over prediction functions with limited observed data.
CNP \cite{garnelo2018conditional} is a special case of the NP family, which encodes data into a deterministic hidden variable that parametrizes the function. As a result, it does not introduce any uncertainties. To address the limitation of CNP, neural process (NP) \cite{garnelo2018neural} is a stochastic process that learns a latent variable to model an underlying distribution over functions, from which we can sample a function for downstream tasks. ANP \cite{kim2018attentive} marries the merits of CNP and NP by incorporating deterministic and stochastic paths in an attentive way. NPs have also been applied to many few-shot problems, such as modeling stochastic physics fields \cite{holderrieth2021equivariant}, node classification \cite{cangea2022message}, recommendation \cite{lin2021task}, and link prediction \cite{liang2022neural,luo2023graph}. However, none of them applies NPs to FKGC. Besides, existing NPs assume underlying functions satisfying a simple Gaussian distribution, which cannot handle the complex distribution of KG completion functions. 

\subsection{Normalizing Flows}
Normalizing flows (NFs) \cite{papamakarios2021normalizing} employ a sequence of bijective mapping functions to transform a simple distribution into a complex target distribution. 
Normalizing flows attract increasing attention from machine learning researchers \cite{kobyzev2020normalizing}. Rezende et al. \cite{rezende2015variational} introduce the Planar and Radial flows, which are relatively simple but easy to compute. They apply these flows to approximate the posterior distribution in the variational inference. Kingma et al. \cite{kingma2016improved} introduce a inverse autoregressive flow, which is more efficient in the reverse process. RealNVP \cite{dinh2016density} uses the affine coupling functions for coupling flows, which is computationally efficient but limited in expressiveness. NFs are also applied to many tasks, such as image generation \cite{kingma2018glow}, machine translation \cite{setiawan2020variational}, and time series analysis \cite{de2020normalizing}. In this paper, we integrate the NF with NP to model a complex distribution of KG completion functions. 
\section{Preliminary and Problem Definition}\label{sec:preliminary}
In this section, we introduce key concepts used in this paper and formally define our problem.

\subsection{Preliminary}

\subsubsection{Neural Process}
Neural process (NPs) \cite{garnelo2018neural,garnelo2018conditional} are stochastic processes that model a distribution over prediction functions $f: X\to Y$. Specifically, the function $f$ is assumed to be parameterized by a high-dimensional random vector $z$. The distribution of functions can be represented by $P(z|C)$, which is empirically assumed to be a Gaussian distribution conditioned on limited \textit{context data} $\mathcal{C}=\{(x_i,y_i)\}_{i=1}^n$.
By sampling a $z$ from the distribution, the NP can readily obtain a function specialized for new prediction tasks. Thus, prediction likelihood on the \textit{target data} $\mathcal{D}=\{(x_i,y_i)\}_{i=n+1}^{n+m}$ is modeled as 
\begin{equation}
        \setlength\abovedisplayskip{1pt}
        \setlength\belowdisplayskip{1pt}
        P(y_{nn+1:n+m}|x_{n+1:n+m},\mathcal{C})=\int_{z} P(y_{n+1:n+m}|x_{n:n+m},z)P(z|\mathcal{C})dz,
\end{equation}
where $n$ and $m$ respectively denote numbers of samples in $\mathcal{C}$ and $\mathcal{D}$, $P(z|\mathcal{C})$ is calculated by an \textit{encoder} using context data, and $P(y_{n+1:n+m}|x_{n:n+m},z)$ is modeled by a \textit{decoder} to realize the function and predict  labels for target data. Since the real distribution of $z$ is intractable, NPs can be trained using the amortized variational inference. The parameters of encoder and decoder are optimized by maximizing the evidence lower bound (ELBO), formulated as
\begin{equation}
    \begin{split}
        &\log P(y_{n+1:n+m}|x_{n:n+m},\mathcal{C})\geq\\
        &\mathbb{E}_{Q_\theta(z|\mathcal{C},\mathcal{D})}[\log P_\phi(y_{n+1:n+m}|x_{n:n+m},z)]-KL\big(Q_\theta(z|\mathcal{C},\mathcal{D})||P_\theta(z|\mathcal{C})\big),
    \end{split}
\end{equation}
where $\theta$ denotes the parameters of encoder, $\phi$ denotes the parameters of decoder, and $Q_\theta(z|\mathcal{C},\mathcal{D})$ approximates the true posterior distribution.

\subsubsection{Normalizing Flows}
Normalizing flows (NFs) \cite{rezende2015variational,papamakarios2021normalizing} enables to transform a simple (e.g., Gaussian) distribution into an expressive complex distribution by applying $T$ steps bijective transformations. First, NFs samples a $z_0\in\mathbb{R}^d$ from the base distribution $Q_0(z_0)$. Following the change of variable rule, there exists an invertible and smooth mapping function $g_i:\mathbb{R}^d\to \mathbb{R}^d$. By stacking a chain of $g_i$ (flows), the resulting variable $z_T$ is given by
\begin{equation}
    \setlength\abovedisplayskip{1pt}
    \setlength\belowdisplayskip{1pt}
    z_T=g_T\circ \ldots \circ g_2 \circ g_1(z_0).
\end{equation}
The final complex distribution of $z_T$ is obtained by
\begin{equation}
    \setlength\abovedisplayskip{1pt}
    \setlength\belowdisplayskip{1pt}    Q_T(z_T)=Q_0(z_0)\prod_{i=1}^T\left|\det\frac{\partial g_i}{\partial z_{i-1}}\right|^{-1},
\end{equation}
where $|\det\frac{\partial g_i}{\partial z_{i-1}}|^{-1}$ is the absolute value of the determinant of the Jacobian of $g_i$ at $z_{i-1}$.

\subsection{Problem Definition}
A Knowledge Graph (KG) is represented as a collection of triples $\mathcal{G}=\{(h,r,t)\subseteq \mathcal{E}\times \mathcal{R} \times \mathcal{E}\}$, where $\mathcal{E}$ and $\mathcal{R}$ respectively denote the set of entities and relations. Few-shot knowledge graph completion (FKGC) aims to predict unseen facts of a given relation with few-shot entity pairs \cite{zhang2020few}.

\noindent\textbf{Definition 1: Few-shot Knowledge Graph Completion (FKGC).}
Given a relation $r\in\mathcal{R}$ and its $K$-shot support set $\{(h_i,r,t_i)\}_{i=1}^K$, we focus on designing a KG completion function $f_r: h\to t$ that predicts possible tail entities $t_q$ in the query set $\{(h_q,r,?)\}$, where $K$ denotes the number of triples in the support set.

In FKGC, the training process is to optimize the model under a set of training relations as well as their corresponding support and query sets, denoted as $\mathcal{R}_{train}$. In the testing phase, the model is required to infer new triples for relations in $\mathcal{R}_{test}$, given a corresponding few-shot support set. The relations in $\mathcal{R}_{test}$ are not seen in the training set, i.e., $\mathcal{R}_{train} \cap \mathcal{R}_{test} = \emptyset$.

To integrate the neural process with FKGC, for each few-shot relation $r$, we treat its support set as context data $\mathcal{C}_r=\{(h_i,r,t_i)\}_{i=1}^K$ and the query set as target data $\mathcal{D}_r=\{(h_q,r,?)\}$, i.e., $\mathcal{T}_r=\{\mathcal{C}_r,\mathcal{D}_r\}$.
\begin{figure*}[t]
    \centering
    \includegraphics[width=.9\textwidth,trim=0 0 0cm 0, clip]{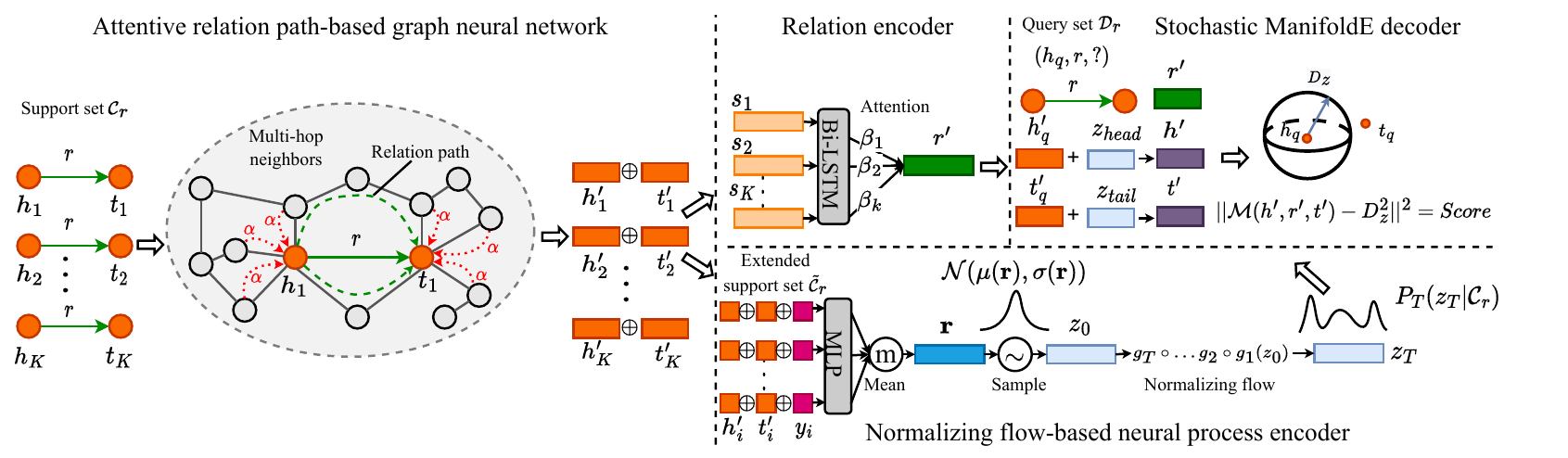}
    \caption{The framework of normalizing flow-based neural process for few-shot knowledge graph completion (\ourmethod).}
    \label{fig:framework}
\end{figure*}

\section{Approach}
\label{sec:approach}
In this section, we introduce our proposed normalizing flow-based neural process (\ourmethod), which consists of four major components: (1) an attentive relation path-based graph neural network to capture the relation path information between entities; (2) a relation encoder to learn the representation of a given few-shot relation; (3) a normalizing flow-based neural process encoder to define a distribution over KG completion functions; (4) a stochastic ManifoldE decoder to predict tail entities. The framework of \ourmethod is illustrated in Fig. \ref{fig:framework}. 

\subsection{Attentive Relation Path-based GNN}
Learning good entity representations is essential for the FKGC task. While entity representations learned from KG embedding methods such as TransE \cite{bordes2013translating} already contain relations information, existing FKGC methods \cite{zhang2020few,sheng2020adaptive} aggregate information from one-hop neighbors to further enhance the entity representations. However, they ignore the \emph{relation path} information between entities, which has shown great power in KG completion \cite{shen2020modeling,wang2021relational}. Intuitively, the breadth-first search (BFS) is the easiest way to obtain paths between entities to facilitate the FKGC \cite{xu2021p}. However, it is time-consuming in large-scale KGs and cannot handle unconnected entity pairs. Recently, many works have adopted the GNN to capture structure and path information in graphs \cite{wang2021relational,you2021identity,luo2021detecting}. Therefore, we propose an attentive relation path-based graph neural network (ARP-GNN) to capture the relation path information between entities. 

Specifically, given an entity $v$, we aggregate its neighbors' information by adopting a attentive relation message passing method. At each layer, we first generate the representation of each neighbor relation $r_i$. Then, we aggregate relation representations attentively to obtain the entity representation. The attentive relation message passing is formulated as follows
\begin{align}
    \setlength\abovedisplayskip{1pt}
    \setlength\belowdisplayskip{1pt}
    m^{l-1} = W_{r_i}(e^{l-1}_v||e^{l-1}_{r_i}||e_{u_i}^{l-1}),(v,r_i,u_i)\in \mathcal{G},\\
    e^{l}_v = \textsf{ReLU}(W^l e^{l-1}_v+ \sum_{r_i\in N(v)}a_{r_i }m^{l-1}_{r_i}+B^l),
\end{align}
where $e_v^{l-1},e^{l-1}_{u_i},e_{r_i}^{l-1}$ denotes the representation of entity $v$, neighbor entity $u_i$ and neighbor relations $r_i$; $||$ denotes the concatenation operation; $W_{r_i}$ denotes a relation-specific linear transformation matrix; $W^l$ and $B^l$ is the weight matrix and bias at layer $l$; $\textsf{ReLU}(\cdot)$ denotes the activation function. To differentiate importance of different relations, the attention weight $a_{r_i}$ is calculated as 
\begin{equation}
    \setlength\abovedisplayskip{1pt}
    \setlength\belowdisplayskip{1pt}
    a_{r_i} = \frac{\textsf{LeakyReLU}\big(W(e^{l-1}_v||m^{l-1}_{r_i})\big)}{\sum_{\tilde{r}\in N(v)\textsf{LeakyReLU}\big(W(e^{l-1}_v||m^{l-1}_{\tilde{r}}\big)}},
\end{equation}
where $W$ is the weight matrix and $\textsf{LeakyReLU}(\cdot)$ is the activation function.

Intuitively, the GNN message passing is essentially a simulation of the BFS, which exhibits the ability to capture paths between entities. By using the GNN, we can excavate the parallel computation ability of GPUs to accelerate the process of path extraction. Besides, the ARP-GNN can also differentiate the contribution of each relation in paths by using an attention mechanism.
In this way, by stacking $L$-layer ARP-GNN, we can embed all the length-$L$ relation paths into the entity representation, which is formulated as 
\begin{equation}
    \setlength\abovedisplayskip{1pt}
    \setlength\belowdisplayskip{1pt}
    e^L = \textsf{L-layer ARP-GNN}(e^0),\label{eq:arpgnn}
\end{equation}
where $e^0$ denotes the initial entity/relation representation obtained from KG embedding methods (e.g., TransE). Empirically, most entities in KG can be connected within 2-length paths \cite{xu2021p}. Thus, we choose $L=2$ in our experiments. For pairs that cannot be connected within 2-length paths, the information from its 2-hop neighbors still could enhance entity representations \cite{wang2021relational}.

After generating the entity representation, each triple in the support set $(h_i,r,t_i)\in \mathcal{C}_r$ can be represented as the concatenation of its head and tail entities' representation, formulated as
\begin{equation}
    \setlength\abovedisplayskip{1pt}
    \setlength\belowdisplayskip{1pt}
    s_i = h'_i || t'_i,\label{eq:support}
\end{equation}
where $h'_i$ and $t'_i$ denote the representations of head and tail entities generated by the ARP-GNN, respectively.

\subsection{Relation Encoder}
In order to generate the few-shot relation representation, we adopt a widely used attentive Bi-LSTM relation encoder to summarize triples in the support set \cite{zhang2020few,niu2021relational}. Given a set of triple representations $\{s_i\}_{i=1}^K$, their hidden states are calculated as
\begin{equation}
    \setlength\abovedisplayskip{1pt}
    \setlength\belowdisplayskip{1pt}
    s'_K,\ldots,s'_1 = \textsf{Bi-LSTM}(s_K,\ldots,s_1).
\end{equation}
Then, we aggregate the hidden states attentively to obtain the final relation representation, which is formulated as
\begin{align}
    o_i &= \textsf{Tanh}(Ws'_i+B),\\
    \beta_i &= \frac{\exp(o_i)}{\sum_{j=1}^K \exp(o_j)},\\
    r' &= \sum_{i=1}^K\beta_i s'_i.
\end{align}
The relation representation $r'$ is used to predict new facts in the query set.

\subsection{Normalizing Flow-based NP Encoder}
To apply the neural process to FKGC, we propose a normalizing flow-based neural process encoder to define a distribution over the KG completion functions. The encoder is composed of two parts: (1) a neural process encoder to learn the distribution $P_0(z_0|\mathcal{C}_r)$; (2) a normalizing flow to transform the simple distribution into a complex target distribution $P_T(z_T|\mathcal{C}_r)$ that parameterizes the stochastic KG completion functions.

The neural process encoder aims to draw a joint distribution based on the context data (support set). Thus, given a context data $\mathcal{C}_r$, it tries to capture connections between $h_i$ and $t_i$ to infer the distribution over the function $f_r$. To reduce estimation bias of the distribution, we generate $n$ negative triples for each triple in $\mathcal{C}_r$ by randomly replacing the tail entity  $\mathcal{C}^-_r=\{(h_i,r,t^-_i)\}_{i=1}^{nK}$. The extended context data is denoted as $\tilde{\mathcal{C}_r}=\mathcal{C}_r\cup \mathcal{C}^-_r$.
For each triple in $\tilde{\mathcal{C}_r}$, we first adopt the ARP-GNN to generate head and tail entity representations $h_i',t_i'$. Then, we feed them together with an indicator $y_i$ into a MLP to generate the hidden representation $c_i$, which is formulated as 
\begin{equation}
    \setlength\abovedisplayskip{1pt}
    \setlength\belowdisplayskip{1pt}
    c_i = \textsf{MLP}\Big(h_i'||t_i'||y_i\Big), y_i=\left\{
        \begin{aligned}
            &1, (h_i,r,t_i) \in \mathcal{C}_r\\
            &0, (h_i,r,t_i) \in \mathcal{C}^-_r
        \end{aligned}
    \right.,
    \label{eq:encoder}
\end{equation}
where $y_i$ denotes whether $(h_i,r,t_i)\in \tilde{\mathcal{C}_r}$ is a negative triple or not. By introducing negative triples, we can estimate the underlying distribution of function $f_r: h\to t$ more accurately.

The representations of all context pairs $\{c_i\in\tilde{\mathcal{C}_r}\}$ are summarized into a single vector $\mathbf{r}$ by an aggregator function $A(\cdot)$ to define the joint distribution, which must satisfy the condition of \textit{permutation-invariant} \cite{oksendal2003stochastic,garnelo2018neural}. Thus, we select a simple average function to generate the global representation $\mathbf{r}$, which is formulated as
\begin{equation}
    \setlength\abovedisplayskip{1pt}
    \setlength\belowdisplayskip{1pt}
    \mathbf{r} = \frac{1}{|\tilde{\mathcal{C}_r}|}\sum_{c_i\in\tilde{\mathcal{C}_r}} c_i.
\end{equation} 

Previous research \cite{garnelo2018conditional,kim2018attentive,norcliffe2020neural} often considers the distribution $P_0(z_0|\mathcal{C}_r)$ as a Gaussian distribution $\mathcal{N}\big(\mu(\mathbf{r}),\sigma(\mathbf{r})\big)$ parameterized by $\mathbf{r}$, which estimates the uncertainty via defining the mean $\mu(\mathbf{r})$ and variance $\sigma(\mathbf{r})$:
\begin{gather}
    x = \textsf{ReLU}\big(\textsf{MLP}(\mathbf{r})\big),\\
    \mu(\mathbf{r}) = \textsf{MLP}(x),\\
    \sigma(\mathbf{r}) = 0.1 + 0.9*\textsf{Sigmoid}\big(\text{MLP}(x)\big).
\end{gather}
However, the Gaussian distribution is not sufficient to model the complex distribution of KG completion functions. Thus, we propose to use the normalizing flow (NF) to transform a simple distribution $P_0(z_0|\mathcal{C}_r)$ into the complex target distribution $P_T(z_T|\mathcal{C}_r)$.

We first sample a $z_0$ from the base distribution $z_0 \sim \mathcal{N}\big(\mu(\mathbf{r}),\sigma(\mathbf{r})\big)$. By applying a sequence of transformations, the final latent variable $z_T$ is given by 
\begin{equation}
    \setlength\abovedisplayskip{1pt}
    \setlength\belowdisplayskip{1pt}
    z_T = g_T\circ\ldots\circ g_1(z_0),
\end{equation}
where $g_i$ can be any bijective transformation function, such as Planar flow \cite{rezende2015variational}, Real NVP flow \cite{dinh2016density}, and Masked Autoregressive flow \cite{papamakarios2017masked}. The latent variable $z_T$ can be considered as one realisation of the function from the corresponding stochastic process modeled by distribution $P_T(z_T|\mathcal{C}_r)$.

The target distribution of $P_T(z_T|\mathcal{C}_r)$ is defined via the change of variable rule as
\begin{equation}
    \setlength\abovedisplayskip{1pt}
    \setlength\belowdisplayskip{1pt}P_T(z_T|\mathcal{C}_r)=P_0(z_0|\mathcal{C}_r)\prod_{i=1}^T\left|\det\frac{\partial g_i}{\partial z_{i-1}}\right|^{-1}.\label{eq:nf}
\end{equation}
Noticeably, the $P_T(z_T|\mathcal{C}_r)$ not only defines the distribution over functions, but also estimates uncertainties. When the support set is limited or noisy, the entropy of $P_T(z_T|\mathcal{C}_r)$ is higher, indicating the model is more uncertain to its predictions. We detailly analyze the uncertainty captured by \ourmethod in section \ref{sec:uncertainty}.

\subsection{Stochastic ManifoldE Decoder}
The decoder aims to predict tail entities for triples in the query set $(h_q,r,?)\in \mathcal{D}_r$. Conventional FKGC methods often adopt a naive dot product \cite{sheng2020adaptive} or a simple score function (e.g., TransE \cite{chen2019meta}) to predict scores of candidate tails, which cannot effectively handle complex relations (e.g., one-to-one and one-to-many relations). By modeling the triples in a manifold sphere, ManifoldE \cite{xiao2016one} alleviates the \textit{ill-posed algebraic system} and \textit{over-strict geometric form} problems, which impair performance when dealing with complex relations. 
Given a triple $(h,r,t)$, ManifoldE tries to model the triple in a manifold sphere, which is formulated as
\begin{align}
    \setlength\abovedisplayskip{1pt}
    \setlength\belowdisplayskip{1pt}
    s(h,r,t) = ||\mathcal{M}(h,r,t)-D^2_r||^2,
\end{align}
where $D_r$ is a relation-specific manifold parameter and $\mathcal{M}(h_q,r,t_q)$ is a manifold function. We use the TransE as the manifold function in experiments.

However, ManifoldE is proposed for scenarios with abundant triples, which cannot be directly applied to few-shot settings. Thus, we propose a stochastic ManifoldE decoder  (SManifoldE), which combines the merit of ManifoldE \cite{xiao2016one} and neural processes to address the few-shot KG completion problem.

For each query triple $(h_q,r,t_q)$, we first use the ARP-GNN to generate the head and tail entities' representations $h'_q,t_q'$ and use the relation encoder to generate the relation representation $r'$. Then, given the $z_T$ calculated by the normalizing flow-based neural process encoder, we adopt two independent MLPs to map them into the space of head and tail entities, which are formulated as
\begin{align}
    \setlength\abovedisplayskip{1pt}
    \setlength\belowdisplayskip{1pt}
    z_{head} = \textsf{MLP}_{head}(z_T),~
    z_{tail} = \textsf{MLP}_{tail}(z_T).
\end{align}
Following that, we randomly project $h'_q,t'_q$ into the hyperplanes defined by $z_T$, which is formulated as
\begin{align}
    \setlength\abovedisplayskip{1pt}
    \setlength\belowdisplayskip{1pt}
    h' = h'_q + z_{head},~
    t' = t'_q + z_{tail}.
\end{align}
In this way, we can make them intersect more easily and improves the prediction accuracy. Finally, the score calculated by stochastic ManifoldE decoder is formulated as
\begin{equation}
    \setlength\abovedisplayskip{1pt}
    \setlength\belowdisplayskip{1pt}
  s_M(h_q,r,t_q) = ||\mathcal{M}(h',r',t')- D^2_z||^2,\label{eq:score}
\end{equation}
where $D_z\in \mathbb{R}^1$ is also obtained by a MLP function $D_z = \text{MLP}(z_T)$.

By incorporating the stochasticity brought by $z_T$ into the prediction results, our stochastic ManifoldE is able to deal with complex relations in few-shot settings.

\subsection{Optimization and Testing}
\noindent\textbf{Training.}
Sampling a training relation $\mathcal{T}_r=\{\mathcal{C}_r,\mathcal{D}_r\}$ from $\mathcal{R}_{train}$, \ourmethod is optimized by maximizing the evidence lower bound (ELBO) to minimize the prediction loss on
target data given the context data: $\log  P(t_q|h_q,r,\mathcal{C}_r)$. The ELBO loss can be derived as 
\begin{align}
    \log P(t_q&|h_q, r, \mathcal{C}_r) = \int_{z}Q(z)\log\frac{P(t_q, z|h_q,r,\mathcal{C}_r)}{P(z|\mathcal{C}_r)},\\
    &=\int_{z}Q(z)\log\frac{P(t_q, z|h_q,\mathcal{C}_r)}{Q(z)} + KL\big(Q(z)||P(z|\mathcal{C}_r)\big),\\
    &\geq \int_{z}Q(z)\log\frac{P(t_q, z|h_q,r,\mathcal{C}_r)}{Q(z)},\\
    &= \mathbb{E}_{Q(z)}\log\frac{P(t_q, z|h_q,r,\mathcal{C}_r)}{Q(z)},\\
    &= \mathbb{E}_{Q(z)}[\log P(t_q|h_q,r,z) + \log\frac{P(z|\mathcal{C}_r)}{Q(z)}],\\
    &= \mathbb{E}_{Q(z)}[\log P(t_q|h_q,r,z)] 
    - KL\big(Q(z)||P(z|\mathcal{C}_r)\big),\label{eq:elbo}
\end{align}
where $Q(z)$ denotes the true posterior distribution of $z$. During training, $Q(z)$ is approximated by $Q_T(z_T|\mathcal{C}_r,\mathcal{D}_r)$ using the neural process encoder and normalizing flow, which is formulated as
\begin{equation}
    \setlength\abovedisplayskip{1pt}
    \setlength\belowdisplayskip{1pt}
    Q(z) \simeq Q_T(z_T|\mathcal{C}_r,\mathcal{D}_r) = Q_0(z_0|\mathcal{C}_r,\mathcal{D}_r)\prod_{i=1}^T\left|\det\frac{\partial g_i}{\partial z_{i-1}}\right|^{-1}.
\end{equation}
With the normalizing flow, the KL divergence does not have a closed form solution. Thus, we rewrite the ELBO loss as
\begin{align}
    \mathcal{L}_{ELBO} &= \mathbb{E}_{Q_0(z_0|\mathcal{C}_r,\mathcal{D}_r)}[\log P(t_q|h_q,r,z_T) - \log \frac{Q_T(z_T|\mathcal{C}_r,\mathcal{D}_r)}{P(z_T|\mathcal{C}_r)}],\\
    \begin{split}
    &= \mathbb{E}_{Q_0(z_0|\mathcal{C}_r,\mathcal{D}_r)}[\log P(t_q|h_q,r,z_T) - \log Q_0(z_0|\mathcal{C}_r,\mathcal{D}_r) \\
    &+ \sum_{i=1}^T\left|\det\frac{\partial g_i}{\partial z_{i-1}}\right|+\log P(z_T|\mathcal{C}_r)],
    \end{split}\label{eq:loss}
\end{align}
where $\mathbb{E}_{Q_0(z_0|\mathcal{C}_r,\mathcal{D}_r)}$ is approximated via Monte-Carlo sampling.

The prediction likelihood $\log P(t_q|h_q,r,z_T)$ is calculated by widely used margin ranking loss, formulated as 
\begin{equation}
    \setlength\abovedisplayskip{1pt}
    \setlength\belowdisplayskip{1pt}
    \log P(t_q|h_q,r,z_T) = -\sum_{q^+,q^-\in \mathcal{D}_r} \max\big(0, S_M(q^-) - S_M(q^+) + \gamma \big),
\end{equation}
where $q^+=(h_q,r,t_q)$ denotes the positive triples in target data, and $q^-=(h_q,r,t^-_q)$ denotes the negative triples by randomly corrupting the tail entity. By using the ranking loss, we try to rank the positive tail entity higher than the negative tails. The training algorithm of \ourmethod is shown in Algorithm \ref{alg:FKGC}.

\noindent\textbf{Testing.}
At testing time, given an unseen relation $r$, we first generate the relation representation $r'$ (Eq. 5-13) and a target distribution $P_T(z_T|\mathcal{C}_r)$ (Eq. 14-19) from its few-shot support set $\mathcal{C}_r$. Then, we sample a global latent variable $z_T$ from $P_T(z_T|\mathcal{C}_r)$. Last, we adopt the stochastic ManifoldE decoder (Eq. \ref{eq:score}) to predict the tail entities for query set $f_r: h_q\to t_q$.

\noindent\textbf{Complexity.} Given a relation with its $K$-shot support set $\mathcal{C}_r$ and $m$ facts in the query set $\mathcal{D}_r$, the complexity of \ourmethod is $O\big((n+1)K+m\big)$, where $n$ denotes the negative sampling size for $\tilde{\mathcal{C}_r}$, since the model only needs to encode each triple in $\tilde{\mathcal{C}_r}$ and predict facts in the query set. This shows that \ourmethod is more efficient as it does not need an additional fine-tuning process required by many meta-learning-based approaches \cite{chen2019meta,niu2021relational}. 

\begin{algorithm}[tbp]
    \caption{The training process of \ourmethod}\label{alg:FKGC}
    \KwIn {Training relations $\mathcal{R}_{train}$; knowledge graph $\mathcal{G}$.}
    \KwOut {Model parameters $\Theta$}
    \While{not done}{
        Sample a relation $\mathcal{T}_r=\{\mathcal{C}_r,\mathcal{D}_r\}$ from $\mathcal{R}_{train}$\;
        Generate a relation representation $r'$ (Eq. 5-13)\;
        Generate a prior distribution $P(z|\mathcal{C}_r)$ using $\mathcal{C}_r$ (Eq. 14-18)\;
        Generate a target base distribution $Q_0(z_0|\mathcal{C}_r,\mathcal{D}_r)$ using $\mathcal{C}_r$ and $\mathcal{D}_r$ (Eq. 14-18)\;
        Sample a $z_0$ from the base distribution $Q_0(z_0|\mathcal{C}_r,\mathcal{D}_r)$ \;
        Apply NF to generate $z_T$ (Eq. 19)\;
        Optimize $\Theta$ using the ELBO loss (Eq. \ref{eq:loss})\; 
    }
\end{algorithm}
\begin{table}[t]
    \centering
    \caption{Statistics of the experimental datasets.}
    \label{tab:dataset}
    \resizebox{\columnwidth}{!}{%
        \begin{tabular}{@{}ccccccc@{}}
            \toprule
            Dataset & \#Relation & \#Entity  & \#Triples & \#Train & \#Valid & \#Test \\
            \midrule
            NELL    & 358        & 68,545    & 181,109   & 51      & 5       & 11     \\
            WIKI    & 822        & 4,838,244 & 5,859,240 & 133     & 16      & 34     \\
            FB15K-237   & 231        & 14,541    & 281,624   & 75      & 11      & 33     \\
            \bottomrule
        \end{tabular}
    }
\end{table}

\begin{table*}[tbp]
    \centering
    \caption{5-shot FKGC results on the NELL, WIKI, and FB15K-237 datasets.}
    \label{tab:results}
    \resizebox{.8\linewidth}{!}{%
        \begin{tabular}{@{}c|cccc|cccc|cccc@{}}
            \toprule
            \multirow{2}{*}{Methods} & \multicolumn{4}{c|}{NELL} & \multicolumn{4}{c|}{WIKI} & \multicolumn{4}{c}{FB15K-237}                                                                                                                                                                                     \\
            \cmidrule{2-13}
                                     & MRR                       & Hits@10                   & Hits@5                    & Hits@1            & MRR               & Hits@10           & Hits@5            & Hits@1            & MRR               & Hits@10           & Hits@5            & Hits@1            \\
            \midrule
            TransE                   & 0.168                     & 0.345                     & 0.186                     & 0.082             & 0.052             & 0.090             & 0.057             & 0.042             & 0.307	&0.537	&0.419	&0.198\\
            TransH                   & 0.279                     & 0.434                     & 0.317                     & 0.162             & 0.095             & 0.177             & 0.092             & 0.047             & 0.284	&0.503	&0.397	&0.181\\
            DisMult                  & 0.214                     & 0.319                     & 0.246                     & 0.140             & 0.008             & 0.134             & 0.078             & 0.035             & 0.237	&0.378	&0.287	&0.164\\
            ComplEx                  & 0.239                     & 0.364                     & 0.253                     & 0.176             & 0.070             & 0.124             & 0.063             & 0.030                          & 0.238	&0.370	&0.281	& 0.169\\
            \midrule
            GMatching                & 0.176                     & 0.294                     & 0.233                     & 0.113             & 0.263             & 0.387             & 0.337             & 0.197             & 0.304             & 0.456             & 0.410             & 0.221             \\
            MetaR                    & 0.261                     & 0.437                     & 0.350                     & 0.168             & 0.221             & 0.302             & 0.264             & 0.178             & 0.403             & 0.647             & 0.551             & 0.279             \\
            FSRL                     & 0.153                     & 0.319                     & 0.212                     & 0.073             & 0.158             & 0.287             & 0.206             & 0.097             & 0.365             & 0.553             & 0.456             & 0.271             \\
            FAAN                     & 0.284                     & 0.451                     & 0.373                     & 0.194             & 0.227             & 0.363             & 0.288             & 0.157             & 0.425             & 0.518             & 0.459             & 0.340             \\
            GANA                     & \underline{0.344}         & \textbf{0.517}            & \underline{0.437}         & \underline{0.246} & \underline{0.351} & \underline{0.446} & \underline{0.407} & \underline{0.299} & \underline{0.458} & \underline{0.656} & \underline{0.575} & \underline{0.349} \\
            \midrule
            \ourmethod               & \textbf{0.460}            & \underline{0.494}         & \textbf{0.471}            & \textbf{0.437}    & \textbf{0.503}    & \textbf{0.668}    & \textbf{0.599}    & \textbf{0.423}    & \textbf{0.538}    & \textbf{0.671}    & \textbf{0.593}    & \textbf{0.476}    \\
            \bottomrule
        \end{tabular}
   }
\end{table*}
\section{Experiment}

\subsection{Datasets and Metrics}
We evaluate the performance of our \ourmethod method against other baselines on three public benchmark datasets: NELL, WIKI, and FB15K-237~\cite{xiong2018one,zhang2020few,niu2021relational,wang2021reform}. Following previous settings \cite{chen2019meta}, relations with more than 50 triples but less than 500 triples are selected as few-shot relations. The first $K$ triples are used as the support set, and the rest is used as the query set. We adopt the public splits \cite{xiong2018one,wang2021reform}, in which the selected relations are divided into 51/5/11 in NELL, 133/16/34 in WIKI, and 75/11/33 in FB15K-237 for training/validation/testing, respectively. The statistics of datasets and their splits are shown in Table \ref{tab:dataset}.

We remove the few-shot relations which are in the original training/validation/testing sets from the knowledge graph $\mathcal{G}$ to prevent information leakage \cite{zhang2020few}. Besides, to fully utilize the knowledge graph, we also sample relations (even not few-shot relations) from $\mathcal{G}$ to enrich the training set, which follows the same setting as previous methods \cite{chen2019meta}.

We adopt two widely used metrics, MRR and Hits@$N$ for a direct comparison with the results reported by previous methods. MRR denotes the mean reciprocal rank of the correct tail entities, and Hits@$N$ denotes the percentage of the correct tail entities in the top-$N$ ranked entities. We set $N$ to 1, 3, and 5 in our experiments, following previous settings.

\subsection{Baselines}\label{sec:baselines}
We compare our \ourmethod method with two groups of baselines: \textbf{Traditional KGC methods.} These methods learn the entity and relation embeddings by modeling the relation structures in KG. We select four models (e.g., TransE \cite{bordes2013translating}, TransH \cite{wang2014knowledge}, DistMult \cite{bordes2014semantic}, and ComplEx \cite{trouillon2016complex}) as the traditional KGC baselines. 
These baselines can be implemented using the open-source code\footnote{\url{https://github.com/thunlp/OpenKE/tree/OpenKE-Tensorflow1.0}}.
\textbf{FKGC methods.} We select five FKGC methods.
including GMatching \cite{xiong2018one}, MetaR \cite{chen2019meta}, FSRL \cite{zhang2020few}, FAAN \cite{sheng2020adaptive}, and GANA \cite{niu2021relational}.
The implementation of these methods can be obtained from the repositories publicized by their authors. In NELL and WIKI, these methods follow the same settings, we directly use results reported by these papers to avoid re-implementation bias. In FB15K-237, we use the code publicized by authors to conduct experiments.



\subsection{Implementation Details}
In experiments, the initial entity/relation embeddings are obtained from the TransE model released by previous research \cite{xiong2018one,wang2021reform}. We choose the Planar flow \cite{rezende2015variational} as the normalizing flow and the score function of TransE as the manifold function in Eq. \ref{eq:score} (i.e., $||h'+r'-t'||^2$). For hyper-parameters, the embedding dimension and $z$ are set to 100 for both NELL as well as FB15K-237, and 50 for WIKI. The ARP-GNN layer $L$ is set to 2. The hidden dimension of Bi-LSTM is set to 700 and its layers are set to 2. The transformation steps of the NF are set to 10. The negative sampling size $n$ is set to 1, and the learning rate is set to 0.001. The margin $\gamma$ and the Monte-Carlo sampling size are both set to 1. The batch size is set to 128 for NELL and FB15K-237, but 64 for WIKI due to the limitation of GPU memory. We use Adam as the optimizer.

\begin{table*}[htbp]
    \centering
    \caption{5-shot FKGC results on the NELL dataset w.r.t. different categories of complex relations.}
    \label{tab:complex}
    \resizebox{0.8\linewidth}{!}{%
        \begin{tabular}{@{}c|cc|cc|cc|cc@{}}
            \toprule
            \multirow{2}{*}{Methods} & \multicolumn{2}{c|}{MRR} & \multicolumn{2}{c|}{Hits@10} & \multicolumn{2}{c|}{Hits@5} & \multicolumn{2}{c}{Hits@1}                                                                                 \\
            \cmidrule{2-9}
                                     & one-to-one                      & one-to-many                          & one-to-one                         & one-to-many                        & one-to-one               & one-to-many               & one-to-one               & one-to-many               \\
            \midrule
            TransE                   & 0.198                    & 0.088                        & 0.397                       & 0.196                      & 0.293             & 0.163             & 0.186             & 0.033             \\
            TransH                   & 0.354                    & 0.136                        & 0.446                       & 0.358                      & 0.497             & 0.320             & 0.304             & 0.053             \\
            DistMult                 & 0.368                    & 0.167                        & 0.476                       & 0.305                      & 0.488             & 0.293             & 0.315             & 0.083             \\
            ComplEx                  & 0.315                    & 0.131                        & 0.347                       & 0.292                      & 0.547             & 0.222             & 0.320             & 0.058             \\
            \midrule
            MetaR                    & 0.334                    & 0.207                        & 0.480                       & 0.393                      & 0.530             & 0.320             & 0.312             & 0.118             \\
            FAAN &0.442	&0.205 &0.559	&0.406 &0.523	&0.337 &0.371	&0.126\\
            GANA                     & \underline{0.499}        & \underline{0.228}            & \underline{0.624}           & \underline{0.424}          & \underline{0.572} & \underline{0.343} & \underline{0.431} & \underline{0.120} \\
            \midrule
            \ourmethod               & \textbf{0.674}           & \textbf{0.447}               & \textbf{0.675}              & \textbf{0.499}             & \textbf{0.675}    & \textbf{0.466}    & \textbf{0.672}    & \textbf{0.413}    \\
            $w/o$ NF                 & 0.644                    & 0.365                        & 0.651                       & 0.425                      & 0.648             & 0.387             & 0.638             & 0.320             \\
            $w/o$ SManifoldE         & 0.665                    & 0.421                        & 0.667                       & 0.456                      & 0.665             & 0.439             & 0.663             & 0.391             \\
            \bottomrule
        \end{tabular}
   }
\end{table*}

\subsection{FKGC Results}
In this section, we report the results of 5-shot FKGC on the NELL, WIKI, and FB15K-237 datasets, which are shown in Table \ref{tab:results}. The best and second-best results are highlighted in bold and underlined, respectively. From the results, we can see that \ourmethod significantly outperforms all baselines and sets new STOA performance on most metrics.

Specifically, \ourmethod outperforms the existing STOA (i.e., GANA) in MRR by 33.7\%, 43.3\%, and 17.5\% on three datasets. This indicates that \ourmethod could rank the correct tail entities much higher in general. Besides, \ourmethod improves the Hits@1 of GANA by 77.6\%, 41.5\%, and 36.4\% on three datasets, which shows that \ourmethod enables to precisely predict the tail entity. Although the Hits@10 of GANA in NELL is slightly higher than us, \ourmethod still beats it in other metrics. The possible reason is that \ourmethod captures relation paths by multi-layer APR-GNN, which is more effective than the 1-hop neighbor information in GANA. Besides, the normalizing flow and SManifoldE enable our method to handle complex relations more effectively.

Traditional KGC methods (e.g., TransE) achieve the worst results, especially on the WIKI dataset, which naturally reflects the fact that they are not designed for few-shot settings. FKGC baselines (e.g., GMatching and FSRL), on the other hand, achieve better performance. Because they design a matching network to capture the similarity between the support set and query set. However, the simple matching network is not expressive enough to capture the complex relations in KG. Thus, their performance is surpassed by some traditional KGC methods that consider complex relations (e.g., TransH, and ComplEx) on NELL.

MetaR adopts the meta-learning framework to update relation representations, and FAAN designs an attention-based neighbor encoder to enhance entity representations, which reaches better results. However, MetaR applies the score function of TransE, and FAAN adopts a simple dot production to predict tail entities, which also fails to handle complex relations. GANA not only considers the neighbor information but also proposes a MTransH score function for complex relations. Thus, GANA outperforms other baselines and achieves the second-best performance.

\subsection{Results on Complex Relations}

To evaluate the effectiveness of \ourmethod on complex relations, we report the results on different categories of complex relations in Table \ref{tab:complex}. We divide the complex relations in NELL into two categories: one-to-one and one-to-many, following the split of GANA \cite{niu2021relational}.

\noindent\textbf{Handling complex relations}.
From the results, we can notice that \ourmethod consistently outperforms other baselines w.r.t. different categories of relations. Although KGC methods that consider complex relations achieve better results in one-to-many relations, they cannot yet adapt to few-shot settings. GANA is the only baseline considering complex relations, which reaches the second-best performance in both one-to-one and one-to-many relations.

\noindent\textbf{Effectiveness of normalizing flow and SManifoldE.}
The ability of handling complex relations can be credited to the normalizing flow (NF) and SManifoldE.
By respectively removing these components, we can see that the performance on one-to-many relations drops significantly. Specifically, the NF defines a more expressive distribution for the stochastic function, which enables the model to handle complex relations complying with different distributions. Meanwhile, the SManifoldE models triples in a manifold sphere to capture complex relations. Besides, by incorporating the neural process, SManifoldE could be more resilient to the few-shot setting.

\subsection{Uncertainty Analysis}\label{sec:uncertainty}
The major advantage of \ourmethod is able to estimate the uncertainty in its predictions. By using the normalizing flow and neural process, we can obtain a distribution of prediction functions $P_T(z_T|\mathcal{C}_r)$ given the support set. The uncertainty of model can be evaluated by the entropy of $z_T$ \cite{naderiparizi2020uncertainty}. The higher the entropy, the more uncertain the model is.

\begin{figure}
    \centering
    \begin{minipage}[b]{.8\columnwidth}
        \centering
    \includegraphics[trim=0 0cm 0cm 0cm,clip,width=1\linewidth]{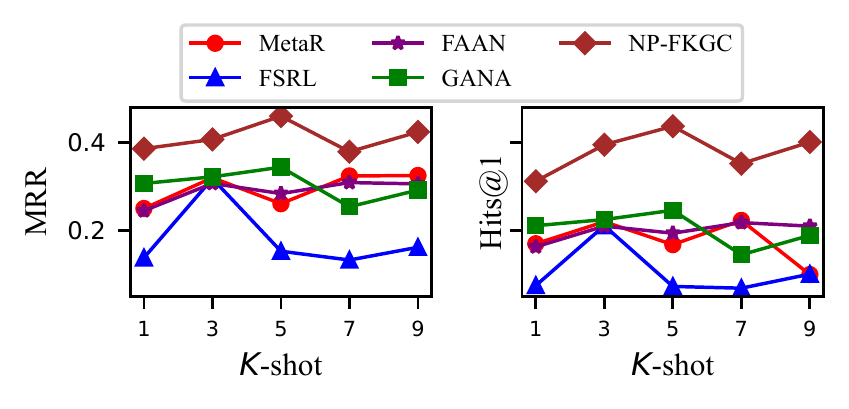}
    \end{minipage}
    \\
    \begin{minipage}[b]{0.43\columnwidth}
        \centering
    \includegraphics[trim=0.cm 0.cm 0.cm 0.0cm,clip,width=1\linewidth]{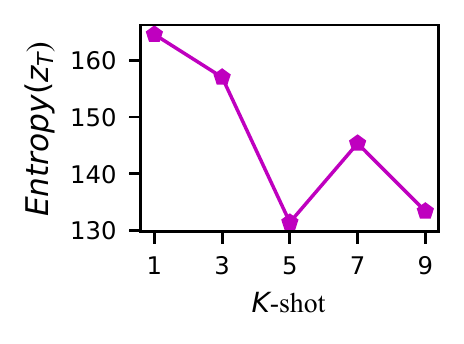}
    \end{minipage}\\
    \begin{minipage}[t]{1\columnwidth}
        \caption{MRR and Hits@1 results (top) and $Entropy(z_T)$ estimated by \ourmethod  (bottom) under different few-shot sizes $K$.}
    \label{fig:kshot}
	\end{minipage}%
\end{figure}

We first evaluate the performance of \ourmethod and other baselines under different few-shot sizes $K$ on the NELL dataset, then illustrate the corresponding $Entropy(z_T)$ estimated by \ourmethod in Fig. \ref{fig:kshot}.
Specifically, we report the MRR and Hits@1 results w.r.t. different few-shot sizes in the top two figures of Fig. \ref{fig:kshot}. From the results, we can see that \ourmethod consistently outperforms other baselines under all $K$ values. The reason is that neural processes could effectively learn a distribution with limited support sets to estimate the underlying facts. 

With the few-shot size increasing (from 1 to 5), neural processes could incorporate new observations to enhance the distribution and predict new facts more accurately. The decrease of $Entropy(z_T)$ shown at the bottom of Fig. \ref{fig:kshot} supports the claim. With more observed data, the $Entropy(z_T)$ decreases, meaning that the model is more certain of its prediction, thus making better predictions.

Noticeably, the performance of \ourmethod drops at $K= 7$. The possible reason is that noise could be introduced into the support set with the increase of $K$ \cite{sheng2020adaptive,wang2021reform}. Existing few-shot methods and neural processes are both sensitive to the quality of the support set \cite{kim2021neural}, which could lead to a performance drop. The increase of $Entropy(z_T)$ at $K=7$ validates the high uncertainty of \ourmethod. When the model is more uncertain of its predictions, the prediction performance drops. In most cases, our \ourmethod can capture the prediction uncertainty, thereby achieving reliable predictions.


\subsection{Ablation Study}
In this section, we conduct ablation studies to analyze the effectiveness of each component in \ourmethod.
In Table \ref{tab:ablation}, we report the 5-shot FKGC results of different model variants on NELL dataset. From the results, we can see that every component could benefit model performance. Specifically, by removing the NP, the model would degenerate to the conventional ManfoldE model with a GNN, which is not effective in few-shot settings. Without the NF and SManfoldE, the model would not be able to handle complex relations. Besides, the performance of the model drops significantly by removing ARP-GNN, which demonstrates the importance of considering the relation path information.

\subsection{Study of Normalizing Flow}
In this section, we further study the impact of normalizing flow in \ourmethod.
First, in Table \ref{tab:flow}, we investigate the performance of \ourmethod with different normalizing flows (e.g., Planar flow \cite{rezende2015variational}, Radial flow \cite{rezende2015variational}, and RealNVP flow \cite{dinh2016density}). From the results, we can see that Planar flow achieves the best performance compared to other flows. This indicates that a simple NF would be enough to model the complex distributions of KG completion function.

Second, we explore the model performance under various flow transformation steps $T$. From Table \ref{tab:flowk}, we can see that the performance of \ourmethod increases with the flow transformation step $T$. The reason is that the NF could model a more complex distribution with increased steps, which also improves the performance of handling complex relations (e.g., one-to-many). However, stacking too many flows would impair the computational speed and cause overfitting. Thus, we set $T$ to 10 to balance the performance and efficiency.

Third, even the NF is helpful to improve the expressiveness of the NP, the computational cost of the NF may get higher as the transformation steps increase. In Table \ref{tab:time}, we provide the running time (second) of each training epoch and the total test time under different flow steps $T$. From the results, we can observe that the time grows slightly as $T$ increases. We can also choose a computationally efficient flow or a relatively small $T$ to further decrease the computation time.

\begin{table}[tbp]
    \centering
    \caption{Ablation study on the NELL dataset.}
    \label{tab:ablation}
    \resizebox{0.7\columnwidth}{!}{%
        \begin{tabular}{@{}ccccc@{}}
            \toprule
            Variants        & MRR            & Hits@10        & Hits@5         & Hits@1         \\ \midrule
            \ourmethod      & \textbf{0.460} & \textbf{0.494} & \textbf{0.471} & \textbf{0.437} \\
            $w/o$ NP        & 0.415          & 0.471          & 0.440          & 0.385          \\
            $w/o$ NF        & 0.364          & 0.405          & 0.377          & 0.333          \\
            $w/o$ SManfoldE & 0.419          & 0.445          & 0.429          & 0.399          \\
            $w/o$ ARP-GNN   & 0.102          & 0.191          & 0.128          & 0.053          \\ \bottomrule
        \end{tabular}%
    }
\end{table}

\begin{table}[tbp]
    \centering
    \caption{Results with different flows on the NELL dataset.}
    \label{tab:flow}
    \resizebox{0.8\columnwidth}{!}{%
        \begin{tabular}{@{}ccccc@{}}
            \toprule
            Methods                 & MRR            & Hits@10        & Hits@5         & Hits@1         \\
            \midrule
            \ourmethod $w/$ Planar  & \textbf{0.460} & \textbf{0.494} & \textbf{0.471} & \textbf{0.437} \\
            \ourmethod $w/$ Radial  & 0.401          & 0.451          & 0.417          & 0.373          \\
            \ourmethod $w/$ RealNVP & 0.449          & 0.500          & 0.457          & 0.424          \\\bottomrule
        \end{tabular}%
    }
\end{table}

\begin{table}[tbp]
    \centering
    \caption{MRR under different flow step $T$ on the NELL dataset.}
    \label{tab:flowk}
    \resizebox{.9\columnwidth}{!}{%
        \begin{tabular}{@{}ccccccccc@{}}
            \toprule
            T   & 2     & 4     & 6     & 8     & 10    & 12    & 14             & 16    \\ \midrule
            All & 0.403 & 0.390 & 0.393 & 0.339 & 0.460 & 0.461 & \textbf{0.472} & 0.395 \\
            one-to-one & 0.644 & 0.604 & 0.634 & 0.642 & 0.674 & 0.648 & \textbf{0.687} & 0.652 \\
            one-to-many & 0.403 & 0.409 & 0.398 & 0.413 & 0.447 & 0.447 & \textbf{0.458} & 0.394 \\\bottomrule
        \end{tabular}%
    }
\end{table}

\begin{table}[tbp]
    \centering
    \caption{Running time (second) under different flow steps  $T$.}
    \label{tab:time}
    \resizebox{1\columnwidth}{!}{%
        \begin{tabular}{@{}ccccccccc@{}}
            \toprule
            T   & 2     & 4     & 6     & 8     & 10    & 12    & 14             & 16    \\ \midrule
            Planar (Train/epoch) &	0.207&	0.2073&	0.2107&	0.2105&	0.2104&	0.2142&	0.2167&	0.2174\\\midrule
Planar (Test Total)&435&	436&	438&	439&	440&	442&	443&	445\\\bottomrule

        \end{tabular}%
    }
\end{table}

\begin{figure}[]
    \centering
    \includegraphics[width=.95\columnwidth]{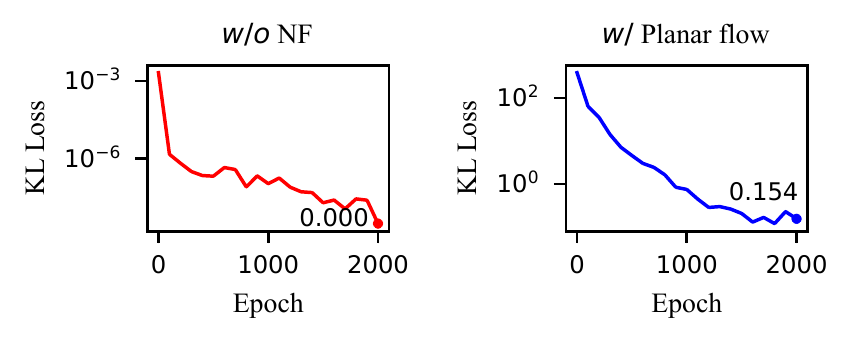}
    \caption{KL loss $w/o$ NF and $w/$ Planar flow.}
    \label{fig:klloss}
\end{figure}

\subsection{Visualization of KL Loss}
Last, we investigate that how the normalizing flow improve the performance by learning a complex distribution. In Fig. \ref{fig:klloss}, we visualize the KL divergence in Eq. \ref{eq:elbo} by epoch for our model $w/o$ NF and $w/$ Planar flow, respectively. From the results, we can see that the KL loss of our model $w/o$ NF starts at a low value (10$^{-3}$) and vanishes to zero. The reason is that the ELBO loss sometimes fails to learn an informative latent variable as indicated by values of the KL term vanishing to 0 \cite{higgins2016beta,alemi2018fixing}, which is also denoted as the \textit{posterior collapse} \cite{razavi2018preventing}. By using normalizing flow, we can achieve a more complex posterior distribution. Optimizing a non-zero KL divergence between posterior and prior distribution would lead the model to learn a meaningful latent variable.  Thus, the KL loss of the model with Planar flow starts at a larger value (10$^{2}$) and decreases to a low but non-zero value (0.154), which also demonstrates the effectiveness of the normalizing flow for alleviating posterior collapse by learning a complex distribution.

\section{Conclusion}
In this paper, we propose a novel normalizing flow-based neural process for few-shot knowledge graph completion. We first integrate a normalizing flow with the neural process to model the complex distribution of KG completion functions and estimate the uncertainty. Then, we design a stochastic ManifoldE decoder to handle the complex relations in few-shot settings. Last, an attentive relation path-based graph neural network is proposed to capture the path information in KGs. Extensive experiments on three benchmark datasets show that our model significantly outperforms the state-of-the-art methods. In the future, we will try to investigate the noise in the support set and further improve the performance.
\section{Acknowledge}
This work is supported by an ARC Future Fellowship (No. FT21010 0097) and is partly based on research sponsored by Air Force Research Laboratory and DARPA under agreement numbers FA8750-19-2-0501 and HR001122C0029. The U.S. Government is authorized to reproduce and distribute reprints for Governmental purposes notwithstanding any copyright notation thereon.
\bibliographystyle{ACM-Reference-Format}
\bibliography{sections/ref.bib}


\end{document}